\documentclass[5p,10pt,twocolumn]{elsarticle}
\usepackage{amssymb,natbib}
\usepackage{graphicx,subfigure,amsmath,color}
\usepackage{epsfig}
\usepackage{multirow}
\usepackage{rotating}
\journal{Physics Letters B}
\begin{document}
\begin{frontmatter}
\title{Shape coexisting intruder states in even Sn isotopes and their influence on the $B_{4/2}<1$ anomaly}
\author[mymainaddress]{P. Buganu}
\author[mymainaddress1]{R. Benjedi}
\author[mymainaddress,mymainaddress2]{R. Budaca\corref{mycorrespondingauthor}}
\cortext[mycorrespondingauthor]{Corresponding author}
\ead{rbudaca@theory.nipne.ro; budacaradu@gmail.com}

\address[mymainaddress]{"Horia Hulubei" National Institute for R\&D in Physics and Nuclear Engineering, Str. Reactorului 30, RO-077125, POB-MG6 Bucharest-M\v{a}gurele, Romania}
\address[mymainaddress1]{High Energy Physics, Astrophysics and Geosciences Laboratory, Department of Physics, Faculty of Science Semlalia, Cadi Ayyad University, P. O. B. 2390, Marrakesh 40000, Morocco}
\address[mymainaddress2]{Academy of Romanian Scientists, 54 Splaiul Independen\c{t}ei, RO-050094, Bucharest, Romania}
\begin{abstract}
The normal and intruder configuration states of the even $^{114-122}$Sn nuclei are described in a geometric model using a double well collective potential. It is shown that subtle changes in the potential shapes can reproduce the mass dependence of the low-lying energy levels and the associated $E0$ and $E2$ transitions. The extracted shape dynamics reveals different instances of shape coexistence with and without mixing, in all considered nuclei. This information is used to track the origin of $B_{4/2}<1$ values reported for $^{114,122}$Sn isotopes, providing thus a phenomenological interpretation.
\end{abstract}
\end{frontmatter}

Having a prominently closed shell at $Z=50$, the tin isotopes form the most extended chain which contains also the doubly magic nuclei $^{100}$Sn and $^{132}$Sn, together with the largest number of stable nuclei. It is therefore a preferred testing ground for the experimental realizations of various nuclear structure models. The description of low-lying states in even Sn nuclei near the stability line, was originally approached from the premise of neutron excitations built on an inert core with $Z=N=50$ \cite{vanGunst}. This hypothesis proved insufficient when applied to excited $0^{+}$ states, expected to behave as pairing phonon excitations \cite{Fielding}. Experimental endeavors of Br\"{o}n {\it et al.} \cite{Bron} provided evidence for the intruder nature of these states originating from $2p$-$2h$ excitations across the proton $Z=50$ shell gap. A systematic analysis of the associated transition probabilities inferred that the excited $0^{+}$ levels constitute rotational band heads with sizable deformation \cite{Backlin}. As a result, the stable Sn nuclei became prime examples of shape coexistence phenomenon \cite{CoHeyde,CoGar,CoBon,CoLeoni}. However, the systematic mass-dependence of the intruder states can be explained only by a concerted mechanism involving also $2p$-$2h$ neutron excitations at the $N=66$ mid-shell \cite{Kaneko}. This in turn maximizes the proton-neutron interaction which generates collectivity \cite{CastenMc}. The origin of enhanced $B(E2,2_{1}^{+}\rightarrow0_{1}^{+})$ values \cite{Beausch} can be in principle explained with this rationale. The dynamical emergence of deformation through configuration mixing is also signaled by the abnormally small values of the $B_{4/2}=B(E2,4_{1}^{+}\rightarrow2_{1}^{+})/B(E2,2_{1}^{+}\rightarrow0_{1}^{+})$ ratios. The mechanisms underlying the so-called $B_{4/2}$ anomaly is currently a focus of intense theoretical \cite{Pan,Zhang,Teng,Bo,Radi,PascuBon,BugBen} and experimental \cite{Zhu,Saigy,Goa,Zanon,Pascu} studies. On the microscopic part, the description of stable tin nuclei is also complicated by the effect of the $N=64$ sub-shell \cite{Backlin,Piller}, and the important variations in pairing conditions \cite{Togashi,Monga,Orce}. 

The plethora of structural aspects, manifesting in even Sn nuclei, converges on a fundamental property, which is the nuclear shape. With this in mind, we propose here to describe the low-lying energy spectrum of even $^{114-122}$Sn isotopes based on a geometric approach with the nuclear shape at its foundation. This is done by considering the Bohr-Mottelson Hamiltonian describing the quadrupole shape dynamics \cite{Bohr1,Bohr2}:
\begin{align}
\hat{H}_{BM}=&-\frac{\hbar^{2}}{2B}\left[\frac{1}{\beta^{4}}\frac{\partial}{\partial\beta}\beta^{4}\frac{\partial}{\partial\beta}+\frac{1}{\beta^{2}\sin3\gamma}\frac{\partial}{\partial\gamma}\sin3\gamma\frac{\partial}{\partial\gamma}\right]\nonumber\\
&+\frac{\hbar^{2}}{8B\beta^{2}}\sum_{k=1}^{3}\frac{\hat{L}_{k}^{2}}{\sin^{2}\left(\gamma-\frac{2\pi}{3}k\right)}+V(\beta,\gamma),
\label{Hamiltonian}
\end{align}
where $B$ is the mass parameter, $\hat{L}_{k}$ are operators of angular momentum projection in the intrinsic reference frame defined in terms of Euler angles $\Omega$, $\beta$ and $\gamma$ are the deformation variables, while $V(\beta,\gamma)$ is the potential energy. In what follows, we consider $\gamma$-unstable collective conditions, which means that the total potential depends only on the axial deformation. This choice is based on previous attempts to describe collectively the Sn isotopes \cite{Backlin,Band}, as well as on the default treatment of their even Cd and Te neighbours \cite{Kern,Hicks,Cd1,MPLA,NoiCd,EuTe}. The collective potential is chosen to be a symmetric polynomial of eighth order
\begin{equation}
V(\beta,\gamma)=V(\beta)=\frac{\hbar^{2}}{2B}\left(\beta^{2}+b_{1}\beta^{4}+b_{2}\beta^{6}+b_{3}\beta^{8}\right).\label{octic}
\end{equation}
It can be easily shown that the unique shape of this octic potential is determined by a set of only three independent parameters $\{b_{1},b_{2},b_{3}\}$ \cite{NoiCd}. The above potential can accommodate a spherical minimum corresponding to normal configuration, simultaneously with a deformed minimum associated with intruder states \cite{Ge}. The later becomes accessible through an excitation energy. 

The solution of the resulted differential equation is of the following form:
\begin{equation}
\Psi_{\xi\alpha\tau LM}(\beta,\gamma,\Omega)=F_{\xi\tau}(\beta)\mathcal{Y}_{\tau\alpha LM}(\gamma\Omega),
\end{equation}
where $\mathcal{Y}$ is a $SO(5)$ spherical harmonic \cite{Y} indexed by the angular momentum $L$, its projection $M$ on the $z$-axis in the laboratory frame of reference, the seniority quantum number $\tau$ \cite{Rakavy}, and by an order $\alpha$ distinguishing $L$ states within a fixed $\tau$ multiplet. The separated $\beta$ wave function is expanded in Bessel functions of the first kind as
\begin{equation}
F_{\xi\tau}(\beta)=\sum_{n=1}^{n_{Max}}A_{\xi n}\frac{\sqrt{2}\beta^{-\frac{3}{2}}J_{\nu}(z_{n}^{\nu}\beta/\beta_{W})}{\beta_{W}J_{\nu+1}(z_{n}^{\nu})},\,\,\nu=\tau+\frac{3}{2},\label{bwf}
\end{equation}
using their associated zeros $z$ \cite{Taseli,Noi1} and is normalized using the $\beta^{4}d\beta$ metric. Here, $n_{Max}$ is the dimension of basis truncation, while $\xi=0,1,...,n_{Max}-1$ indexes the diagonalization solutions. Moreover, the expansion is made in respect to a value $\beta_{W}$, which sets the position of the infinite wall simulating the $V(\beta)\rightarrow\infty$ part of the octic potential (\ref{octic}). It is fixed by constraining the corresponding infinite wall to intersect sufficiently tangentially the tail of the octic potential, and is therefore completely determined by its parametrization. 

\begin{figure}[ht!]
\begin{center}
\includegraphics[width=0.48\textwidth]{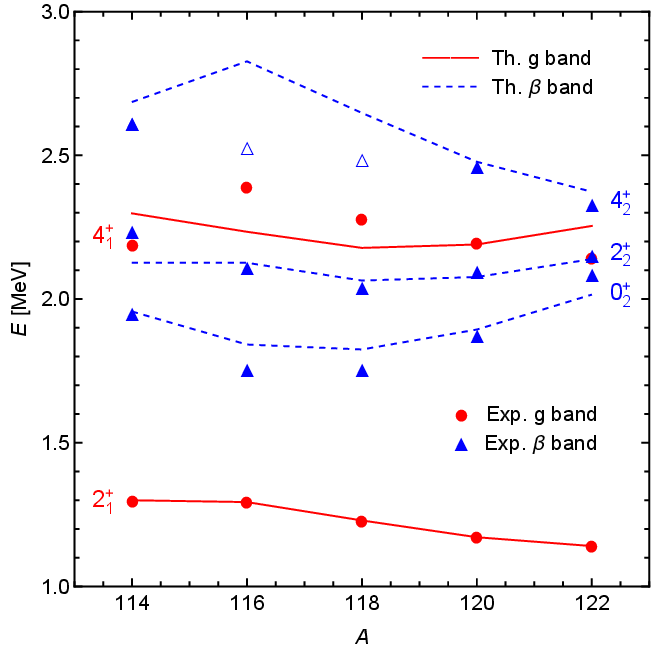}
\end{center}
\vspace{-0.1cm}
\caption{Comparison of experimental \cite{114Sn,116Sn,118Sn,120Sn,122Sn} and theoretical energy levels for the ground and $\beta$ excited band states with $L\le4$ of the even $^{114-122}$Sn isotopes. Open symbols denote data not included in the fitting procedure.}
\label{Levels}
\end{figure}

The model energies are therefore determined only by three independent parameters $\{b_{1},b_{2},b_{3}\}$, up to a scale factor $\hbar^{2}/2B$. The parameters are fixed here by fitting the $4^{+}_{1}$, $0^{+}_{2}$, $2^{+}_{2}$ and in some cases the $4_{2}^{+}$ energy levels, normalized to the corresponding energy of the $2_{1}^{+}$ state, for the $^{114-122}$Sn nuclei. The inclusion of the $4_{2}^{+}$ states is decided when there is a gain in overall agreement with experimental energy levels. The chosen states correspond to $L=2\tau$, and therefore one will use in what follows predominantly the angular momentum designation of states. The scale $\hbar^{2}/2B$ is recovered by equating the theoretical and experimental excitation energy of $2_{1}^{+}$ state, used in normalization. The fitted and scaling parameters are listed in Table \ref{tab1}, while the comparison of experimental and calculated energy levels is shown in Fig. \ref{Levels}. The agreement with experimental energy levels is quiet good, with rms deviation of 15-103 keV for the fitted states. The $\beta$ excited states are especially well reproduced, following therefore the experimental evolution with mass, which is parabolic for $0_{2}^{+}$, slowly varying for $2_{2}^{+}$, and decreasing for $4_{2}^{+}$ states. It is interesting to note that the lowest $0_{2}^{+}$ and $2_{2}^{+}$ states are reported for the $^{116,118}$Sn nuclei, which are above the sub-shell $N=64$ and mid-shell $N=66$.

\setlength{\tabcolsep}{3.8pt}
\begin{table}[th!]
\caption{Model parameters of the octic potential fitted from the low-lying energy levels of the $^{114-122}$Sn isotopes, with the corresponding energy and transition scaling factors  The resulted boundary values are $\beta_W=f(b_{1},b_{2},b_{3})=3.5,3.2,3.3,3.5,3.6$.}
\begin{center}
{\small
\begin{tabular}{cccc|cc}
\hline
&\multicolumn{3}{c|}{Free parameters}&\multicolumn{2}{c}{Scaling factors}\\
\cline{2-4}\cline{5-6}
\\[-0.8em]
Nucleus&$b_{1}$&$b_{2}$&$b_{3}$&$\frac{\hbar^{2}}{2B}$ [keV]&$\beta_{M}$\\
\\[-0.8em]
\cline{2-4}\cline{5-6}
\\[-0.8em]
$^{114}$Sn&5.970466&-1.545435&0.100555&228.28&0.11399\\
$^{116}$Sn&5.207776&-1.444006&0.101072&250.36&0.10888\\
$^{118}$Sn&5.434041&-1.481882&0.102052&230.58&0.10409\\
$^{120}$Sn&6.042888&-1.600728&0.107089&206.24&0.11425\\
$^{122}$Sn&6.577766&-1.620832&0.100572&189.23&0.11376\\
\hline
\end{tabular}}
\end{center}
\label{tab1}
\end{table}

\begin{figure}[ht!]
\begin{center}
\includegraphics[width=0.48\textwidth]{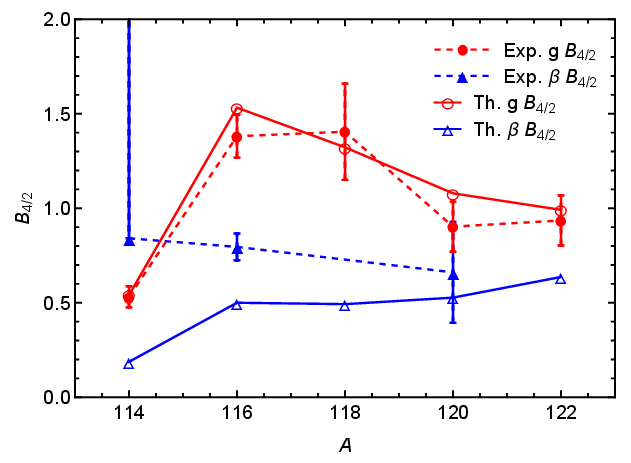}
\end{center}
\vspace{-0.1cm}
\caption{Theoretical and experimental $B_{4/2}$ ratios of the ground and excited bands of $^{114-122}$Sn nuclei, as a function of mass number $A$. Data are collected from Refs. \cite{122Sn,114t,116t,118t,120t}.}
\label{B42}
\end{figure}

Having the model parameters determined, one can calculate transitions probabilities. Of special interest are $E2$ transitions, whose rates are calculated as follows
\begin{eqnarray}
&&B(E2;\xi\tau L\rightarrow\xi'\tau' L')=\left(\frac{3R^{2}Ze}{4\pi}\right)^{2}\beta_{M}^{2}\times\nonumber\\
&&(\tau'\alpha'L';112||\tau\alpha L)^{2}\langle\tau|||Q|||\tau'\rangle^{2}(B_{\xi\tau;\xi'\tau'}^{1})^{2},
\label{BE2}
\end{eqnarray}
where, $(\tau_{1}\alpha_{1}L_{1};\tau_{2}\alpha_{2}L_{2}||\tau_{3}\alpha_{3}L_{3})$ is the $SO(5)$ Clebsch-Gordan coefficient \cite{Y} and
\begin{equation}
\langle\tau|||Q|||\tau'\rangle=\sqrt{\frac{\tau}{2\tau+3}}\delta_{\tau,\tau'+1}+\sqrt{\frac{\tau+3}{2\tau+3}}\delta_{\tau,\tau'-1},
\label{tauop}
\end{equation}
is the $SO(5)$ reduced matrix element of the quadrupole moment \cite{Rowe2,Rowe3}, while $B_{\xi\tau,\xi'\tau'}^{m}$ is the matrix element of $\beta^{m}$ on the $\beta$ wave functions (\ref{bwf}). The units are given by gathering the proton number $Z$, elementary charge $e$, nuclear radius $R=1.2\cdot A^{1/3}$ with $A$ being the mass number, as well as a scaling factor $\beta_{M}$. The latter makes the connection $\beta_{2}=\beta_{M}\beta$ between the geometric variable $\beta$, defining only the nuclear shape, with its microscopic counterpart $\beta_{2}$ which contains also the physical extension of the nuclear mass distribution.

In a similar manner, one can also calculate the monopole transition between the first excited $0_{2}^{+}$ state and the ground state $0_{1}^{+}$,
\begin{equation}
\rho^{2}(E0;0_{2}^{+}\rightarrow0_{1}^{+})=\left(\frac{3Z}{4\pi}\right)^{2}\beta_{M}^{4}\left(B_{20,10}^{2}\right)^{2},
\label{monopole}
\end{equation} 
which is an important observable for shape coexistence and mixing identification.

Note that ratios of transition probabilities of the same kind, do not depend on the scale $\beta_{M}$, and are completely determined by the free parameters $b_{1,2,3}$ deduced from fitting energy levels. This aspect actually constitutes a strong test for the model's performance. This is for example demonstrated in Fig. \ref{B42}, where the theoretical values of intra-band transitions ratio
\begin{equation}
B_{4/2}=\frac{B(E2;4^{+}\rightarrow2^{+})}{B(E2;2^{+}\rightarrow0^{+})}
\end{equation}
are compared with corresponding experimental values. The good agreement with experiment, supports the correct deduction of the octic potential shape. As can be seen from Fig. \ref{B42}, the model calculations also describe quiet well the evolution with mass of the ground band transition ratios. A special attention deserves the fact that the present model can reproduce very small and even sub-unitary ground band $B_{4/2}$ experimental values of the $^{114,120,122}$Sn isotopes.   

\begin{figure*}[ht!]
\begin{center}
\includegraphics[width=0.75\textwidth]{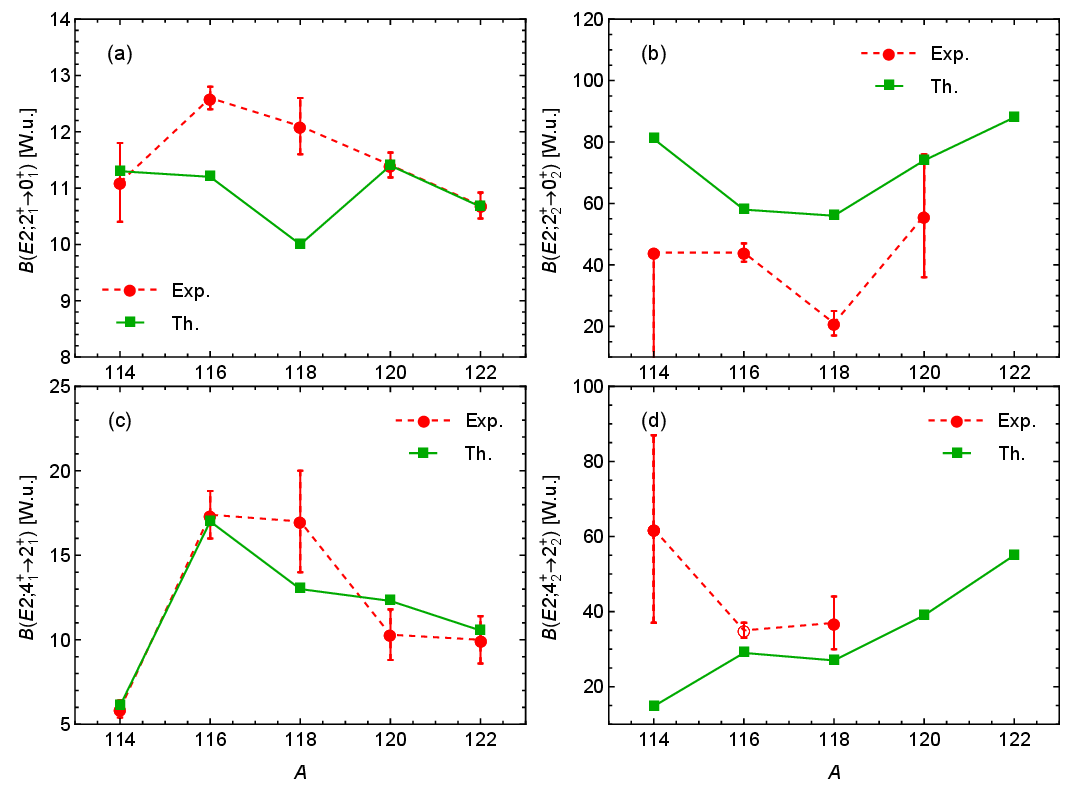}
\end{center}
\vspace{-0.1cm}
\caption{Theoretical and experimental intra-band transition probabilities of the $^{114-122}$Sn nuclei, as a function of mass number $A$. Data are collected from Refs. \cite{122Sn,114t,116t,118t,120t}. Open symbols denote data points which include experimental states not included in the fitting procedure.}
\label{intra}
\end{figure*}

\begin{figure*}[ht!]
\begin{center}
\includegraphics[width=0.75\textwidth]{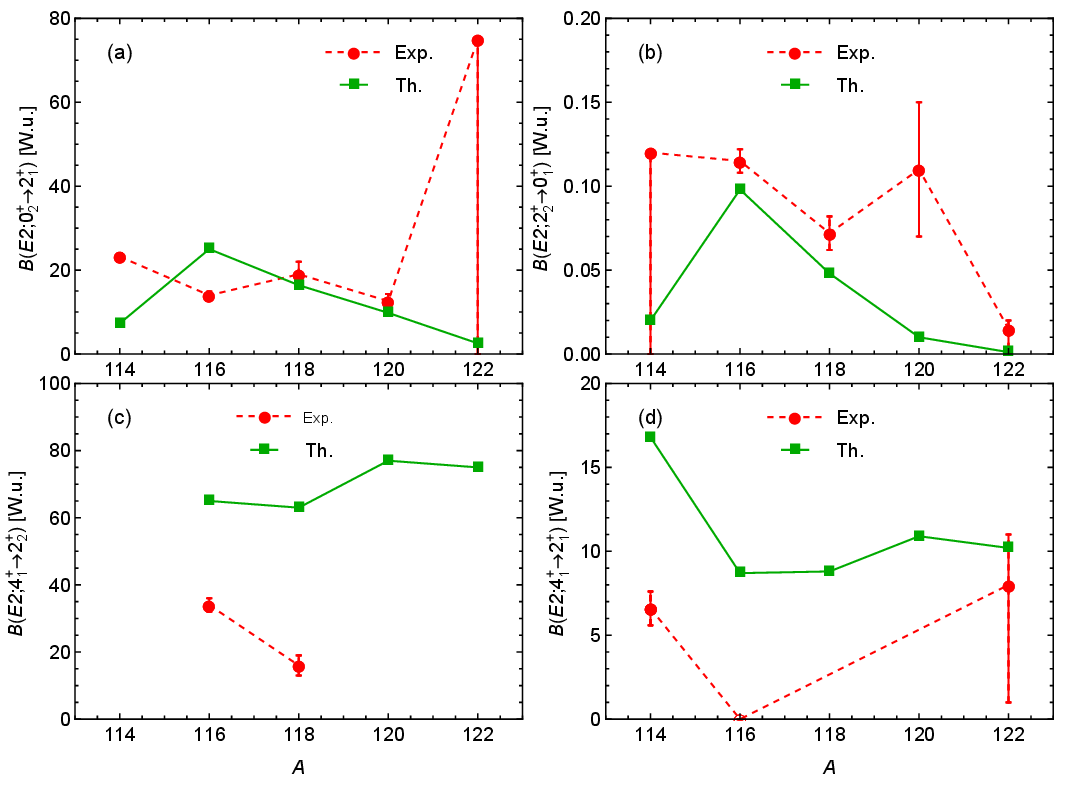}
\end{center}
\vspace{-0.1cm}
\caption{Theoretical and experimental inter-band transition probabilities of the $^{114-122}$Sn nuclei, as a function of mass number $A$. Data are collected from Refs. \cite{122Sn,114t,116t,118t,120t}. Open symbols denote data points which include experimental states not included in the fitting procedure.}
\label{inter}
\end{figure*}

\begin{figure}[ht!]
\begin{center}
\includegraphics[width=0.48\textwidth]{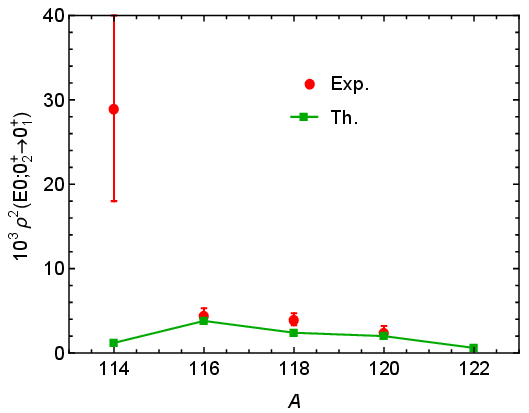}
\end{center}
\vspace{-0.1cm}
\caption{Theoretical and experimental \cite{Kibedi} monopole transition strength for $^{114-120}$Sn nuclei, as a function of mass number $A$.}
\label{mono}
\end{figure}

To have predictions for absolute values of transition probabilities, one determined the scale $\beta_{M}$ by fitting the available experimental data which connect states fitted in the first step. Using minimal errors as weights, one were able to include also the monopole transition strength in this fit. The resulted scaling factors are listed in the last column of Table \ref{tab1}. 

The theoretical predictions for intra-band and inter-band absolute $E2$ transitions are shown in Figs. \ref{intra} and \ref{inter}. A similar comparison is visualized in Fig. \ref{mono}, for the monopole transition between the $0_{1,2}^{+}$ states. The overall agreement with data is very good. Although there are some instances of large discrepancies, the theoretical value is at least of the same order with the available data. The chosen depiction, as a function of mass number, shows that the evolution of experimental data is well reproduced in small isotopic sequences or even along the whole considered chain. For example, theory correctly predicts the minimum and the maximum of the $2_{2}^{+}\rightarrow0_{2}^{+}$ and respectively $4_{1}^{+}\rightarrow2_{1}^{+}$ transitions at the $^{116,118}$Sn isotopes. The departure from experimental $B(E2;2_{1}^{+}\rightarrow0_{1}^{+})$ values of lighter isotopes is in the spirit of diverging measurements within various experiments \cite{Beausch}, which on the other hand converge on the decreasing trend in the two heaviest isotopes. Also, apart from the $^{114}$Sn nucleus, which exhibits a very large measured value, the monopole transition strength is described with impressive fidelity. 

A very good agreement with experiment is also observed in Fig. \ref{inter} (a) and (b), for the inter-band transitions connecting the $0_{1,2}^{+}$ and $2_{1,2}^{+}$ states, whose evolution with mass is well matched by the present calculations. Whereas, the predictions for the inter-band transitions from $4_{1,2}^{+}$ states in general overestimates the few available data points. This aspect comes as a confirmation of the uncertain band assignment of the observed $4_{2}$ state in some of the considered nuclei, where it was not included in the fitting procedure. Due to the adopted $SO(5)$ symmetry, the transition $2_{2}^{+}\rightarrow2_{1}^{+}$ is forbidden in the present model. This is consistent with the small $B(E2)$ values $(\le8)$, 7.8(5), 7.2(10), and 6(2) W.u. reported for $^{114-120}$Sn isotopes \cite{114t,116t,118t,120t}. Whereas the moderate value of 19(3) W.u., measured for the $^{122}$Sn nucleus \cite{122Sn}, points to a more stabilized $\gamma$ deformation.

\begin{figure*}[ht!]
\begin{center}
\includegraphics[width=0.95\textwidth]{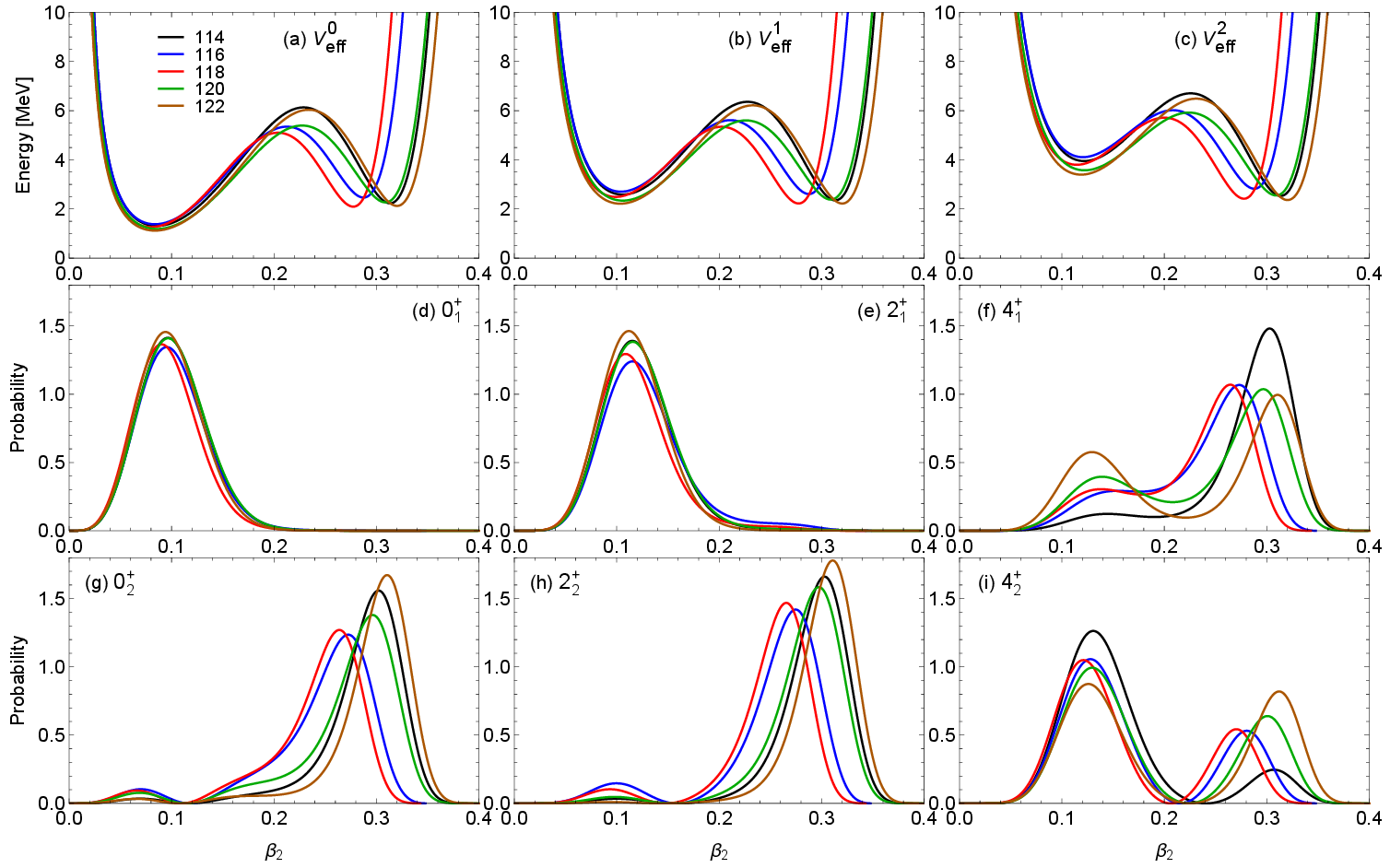}
\end{center}
\vspace{-0.1cm}
\caption{The effective potentials defined by Eq. (\ref{Veff}) for $\tau=1,2,3$ (a,b,c) and the probability distribution (\ref{density}) for the ground states $0_{1}^{+}$, $2_{1}^{+}$, $4_{1}^{+}$ (d,e,f), and $\beta$ excited states $0_{2}^{+}$, $2_{2}^{+}$, $4_{2}^{+}$ (g,h,i), as a function of $\beta_{2}=\beta_{M}\beta$.}
\label{potdens}
\end{figure*}

With the help of the two scaling factors, one can cast the model in accessible energy units and as a function of the traditional axial quadrupole deformation $\beta_{2}$. The dynamics underlying theoretical results can then be ascertained from the state dependent effective potential
\begin{align}
V_{eff}^{\tau}(\beta_{2})=&\frac{\hbar^{2}}{2B}\left[(\tau+2)(\tau+1)\left(\frac{\beta_{M}}{\beta_{2}}\right)^{2}+\left(\frac{\beta_{2}}{\beta_{M}}\right)^{2}\right.\label{Veff}\\
&\left.+b_{1}\left(\frac{\beta_{2}}{\beta_{M}}\right)^{4}+b_{2}\left(\frac{\beta_{2}}{\beta_{M}}\right)^{6}+b_{3}\left(\frac{\beta_{2}}{\beta_{M}}\right)^{8}\right],\nonumber
\end{align}
which includes the centrifugal contribution, when one transfers the problem in the $d\beta$ integration metric. A complementary source of dynamical insights is the corresponding density of deformation probability distribution 
\begin{equation}
\rho_{\xi,\tau}(\beta_{2})=[F_{\xi,\tau}(\beta_{2}/\beta_{M})]^{2}\left(\frac{\beta_{2}}{\beta_{M}}\right)^{4}.
\label{density}
\end{equation}

The effective potentials obtained here for the $^{114-122}$Sn nuclei and shown in Fig. \ref{potdens} (a,b,c), are very similar. The lightest and the heaviest considered isotopes have the deformed minimum at higher deformation. The $^{120}$Sn nucleus retains this high deformation minimum, having however a smaller separating barrier, which is at the same height as for the $^{116,118}$Sn nuclei. The latter two nuclei also have a smaller deformation for the second minimum. Basically, one observes two distinct potential shapes for $^{114,122}$Sn and $^{116,118}$Sn, with $^{120}$Sn nucleus exhibiting a transitional role. This aspect is consistent with the regularities seen in the evolution of energy levels with mass number.     

For all Sn isotopes, the potential minimum at small deformation is deeper in the ground state, then becomes almost degenerate with the well deformed minimum at $\tau=1$ $(L=2)$, and eventually energetically higher for $\tau=2$ $(L=4)$. This is reflected in the spin evolution of the wave function in the ground and $\beta$ excited bands. From Fig. \ref{potdens} (d,e,f), one can observe that the ground state has a well defined near-spherical configuration. As spin is increased to $L=2$, a small amount of the probability distribution spills over to the well deformed configuration associated with the second potential minimum. Nevertheless, the $2_{1}^{+}$ state retains a predominantly spherical deformation. The situation is drastically changed for the $4_{1}^{+}$ state, whose probability distribution is spread or split in two comparable peaks positioned over the corresponding effective potential minima. This corresponds to the shape coexistence scenario, where the $4_{2}^{+}$ state is a superposition of two deformation configurations. The only difference between the considered nuclei, resides in the amount of mixing and separation between the two configurations constituting the $4_{1}^{+}$ state. For the $^{114}$Sn nucleus, one has the most of probability centered on the high deformation configuration. This suggests that this particular nucleus transitions between the $2_{1}^{+}$ and $4_{1}^{+}$ states, from near spherical shape to a well deformed one. As a matter of fact, the disparity between the nature of the two states, leads to a diminished overlap between the corresponding wave functions. This in turn, explains the $B_{4/2}<1$ value reported for this nucleus. On the other extreme, the $4_{1}^{+}$ state in $^{120,122}$Sn nuclei, has the most well defined and separated two-peaks probability distribution. This is another mechanism for a milder destructive overlap with the lower $2_{1}^{+}$ state, which leads to a small $B_{4/2}\approx1$ ratio. The structure of the $4_{1}^{+}$ state, once again shows the particularity of the $^{116,118}$Sn nuclei, whose distinct deformation configurations are closer to each other, merging into a single probability peak with a medium bulge on its less deformed side. This behavior is in agreement with the large shape fluctuations well known to occur in shape phase transition regions \cite{Casten,Prasek}.   

The wave function probability distribution of the $\beta$ excited band, shown in Fig. \ref{potdens} (g,h,i), is an approximate mirror correspondence to the ground band. More precisely, the $0_{2}^{+}$ and $2_{2}^{+}$ states strongly favor a well deformed configuration, even though they have a node. This situation corresponds to the phenomenon of shape coexistence without mixing, where the ground band $L=0,2$ states have a near spherical structure, while the corresponding excited band states exhibit well separated deformed configurations. The node in these wave functions coincides with the position of the ground band state peak, and separates a very small amount of probability at small deformation. In the case of the $0_{2}^{+}$ state, the shape of the high deformation peak in conjunction with the separated low deformation one, suggests a coexistence between a low probability vibration with turning points still in the small deformation effective potential well, and a separate rotational configuration constructed on the high deformation minimum of the same effective potential. A similar three configuration structure of $\beta$ excited states was identified within a related approach also for $^{94}$Zr \cite{BugBen}, $^{96}$Mo \cite{NoiMo,Noi96} and $^{42,44}$Ca \cite{NoiCa} nuclei. The intermediate turning point disappears in the next $2_{2}^{+}$ state. The vibrational nature of the $4_{2}^{+}$ state is more pronounced. Its high energy facilitates the oscillations between the two separate deformation regimes. Note that this time the small deformation configuration is favored. Its opposition to the structure of the lower $2_{2}^{+}$ state, makes its decay weaker, a fact reflected in very small $B_{4/2}$ values for this excited band. 

It is interesting to note that the opposing deformation of the $0^{+}$ and $2^{+}$ states in the ground and $\beta$ excited bands, as well as the mixed nature of the $4_{1,2}^{+}$ states, is also reported for $^{116}$Sn in Ref. \cite{Kaneko}, with the difference that the driving deformation is triaxial. 

In conclusion, we have described with good accuracy the evolution with mass of the low-lying energy levels and the associated electric transitions in even $^{114-122}$Sn nuclei, using a collective geometrical model employing a double well octic potential. The obtained degree of fidelity to the experimental data, allowed us to draw some conclusions regarding the shape dynamics of the considered nuclei. It is thus confirmed that $L=0,2$ states of the ground band are near spherical, whereas those from the excited $\beta$ band are well deformed. The mid-shell $^{116-118}$Sn nuclei exhibit a smaller deformation in the excited band than their neighbors. The considered nuclei become more distinguished through the structure of the $4_{1}^{+}$ state, which displays a shape coexisting character with different degrees of mixing. The small mixing of this state, obtained for the $^{114}$Sn nucleus, confers it a predominantly large deformation in contrast to its lower spin band members. This change in structure explains the $B_{4/2}<1$ anomaly reported for this nucleus. A similar, but milder inhibiting mechanism for the $4_{1}^{+}\rightarrow2_{1}^{+}$ transition, is observed also for the $^{120,122}$Sn nuclei, which exhibit a $B_{4/2}\approx1$ ratio. In this case, the destructive overlap with the $2_{1}^{+}$ state, is due to the almost equal superposition of well separated small and large deformation configurations of the $4_{1}^{+}$ state. We therefore conclude that the small $B_{4/2}$ values are due to sudden change in the structure between $2^{+}$ and $4^{+}$ states, which must contain distinct deformation configurations.

\section*{Acknowledgments}
This work was supported by grants of the Ministry of Research, Innovation and Digitization, CNCS - UEFIS-CDI, Project number PN-IV-P1-PCE-2023-0273, within PNCDI IV, and Project number PN-23-21-01-01/2023.

\end{document}